# Universal correlations between the fragility and interparticle repulsion of glass-forming liquids

Peter Lunkenheimer[1,a], Felix Humann[1], Alois Loidl[1], and Konrad Samwer[2]

**AFFILIATIONS**

[1]Experimental Physics V, Center for Electronic Correlations and Magnetism, University of Augsburg, 86159 Augsburg, Germany

[2]I. Physikalisches Institut, Universität Göttingen, 37077 Göttingen, Germany

[a] **Author to whom correspondence should be addressed:** peter.lunkenheimer@physik.uni-augsburg.de

**Abstract**

A recently published analytical model, describing and predicting elasticity, viscosity, and fragility of metallic melts, is applied for the analysis of about 30 nonmetallic glassy systems, ranging from oxide network glasses to alcohols, low-molecular-weight liquids, polymers, plastic crystals, and even ionic glass formers. The model is based on the power-law exponent $\lambda$ representing the steepness parameter of the repulsive part of the inter-atomic or -molecular potential and the thermal-expansion parameter $\alpha_T$ determined by the attractive anharmonic part of the effective interaction. It allows fitting the typical super-Arrhenius temperature variation of the viscosity or dielectric relaxation time for various classes of glass-forming matter, over many decades. We discuss the relation of the model parameters found for all these different glass-forming systems to the fragility parameter $m$ and detect a correlation of $\lambda$ and $m$ for the non-metallic glass formers, in accord with the model predictions. Within the framework of this model, thus the fragility of glass formers can be traced back to microscopic model parameters characterizing the intermolecular interactions.

**I. INTRODUCTION**

Despite its outstanding importance in many areas of modern technology as well as in our daily life, the process of glassy freezing is hardly understood and controversially debated. Especially, to understand and model the enormous non-canonical variation of viscosity over many orders of magnitude in a rather narrow temperature range, which is typical for glass-forming matter, still is a great challenge.[1] From a large number of detailed experiments on the temperature dependence of the viscosity or the molecular relaxation dynamics for a variety of different glass formers, it was concluded that vitrification is not at all a simple kinetic phenomenon, but a cooperative slowing down of molecular dynamics. Two limiting cases were identified: (i) ideal Arrhenius behavior $\eta \propto \exp(\Delta/T)$, where $\Delta$ corresponds to an energy barrier in Kelvin or (ii) super-exponential temperature dependence, which can, e.g., be approximated by the Vogel-Fulcher-Tammann (VFT) law.[2,3,4] In the latter case, the viscosity is given by $\eta \propto \exp[B/(T-T_{VF})]$, involving the characteristic Vogel-Fulcher temperature $T_{VF}$. In case (i) the viscosity diverges at $T = 0$ K, while in case (ii) it diverges at $T = T_{VF}$, which may be regarded as the critical temperature of some kind of hidden phase transition, well below the glass-transition temperature $T_g$. This possible transition, however, is inaccessible because any glass former inevitably falls out of thermodynamic equilibrium when cooling it below $T_g$. Later on, Angell and coworkers[5,6,7] proposed to classify glass formers according to their degree of deviation from Arrhenius behavior, defining "strong" and "fragile" liquids as those exhibiting weak or strong deviations, respectively. These deviations can be quantified by defining a fragility index $m$, which corresponds to the slope at $T_g$ of the temperature-dependent viscosity or relaxation time $\tau$ in the Angell plot, $\eta$ (or $\tau$) vs. $T_g/T$.[8,9] A value of $m \sim 16$ corresponds to the strongest glass formers, while the upper limit of fragility is $m \sim 170$.[10]

To determine the temperature dependence of the viscosity, numerous experimental methods have been applied.[1] In non-metallic supercooled liquids, dielectric spectroscopy is of outstanding importance to determine molecular relaxation times over extremely broad temperature and frequency regimes.[11,12,13] Using this technique, the slowing down of molecular motion when approaching the glass transition under cooling can be recorded over more than 18 decades of frequency.[11,13] Often the temperature dependence of the mean molecular relaxation times, as determined in a variety of experimental techniques, scales rather well with the viscosity.[13,14] In the past, the VFT relation often has been successfully used to fit both viscosity and relaxation-time data in the range from the glass transition $T_g$ up to high temperatures, deep in the low-viscosity liquid regime.[6,11,13,15] However, this formula by no means is the only possibility to parameterize such broadband data and it was critically commented that there is no real compelling experimental evidence for the divergence of $\eta(T)$ or $\tau(T)$ below $T_g$, suggested by the VFT law.[16] On the other hand, the existence of a diverging length scale and a concomitant finite critical temperature in glass-forming liquids was deduced from the analysis of third and fifth order dipolar non-linear susceptibilities.[17]

Depending on the system under investigation or on the range of data available for fitting, numerous relations have been suggested as alternatives to the VFT law and a variety of theoretical arguments has been developed to support one or the other of these relations. An early overview of possible parameterizations of the temperature dependence of viscosity or mean relaxation time was given by Angell et al.[18] Further approaches were developed by Cohen and Grest,[19] Kivelson et al.,[20] Souletie and Bertrand,[21] Mirigian and Schweizer,[22,23] Schmidtke et al.,[24,25] and by Mauro and coworkers,[26] to name a few. For example, in the latter model the relaxation times or the viscosities diverge only in the zero-temperature limit. Mauro et al.[26] started from thermodynamic considerations using the Adam-Gibbs equation for entropy[27] to formulate an analytical model based on three parameters, which seems to fit many glass-forming liquids quite well. Despite the success of the Mauro model for certain materials,[15,26] the basic understanding of the Adam-Gibbs-relation connecting viscosity and entropy is still a matter of controversy. Another approach to understand temperature-dependent viscosity changes on a microscopic basis was suggested by Aramov-Milchev, who used an atomic hopping picture to derive an alternative three-parameter fit.[28] A critical comparison of the qualities of some of the different parameterizations mentioned above can be found, e.g., in Refs. 15,29,30,31.

In 2015, Krausser, Samwer, and Zaccone (KSZ) presented a very different ansatz, where the high frequency (affine) shear modulus of the glass-forming liquid is proposed to be the driving parameter of the viscosity.[32] Using the shoving model[33,34] KSZ employed the Born-Huang-ansatz for crystals[35] to derive the shear modulus and its temperature dependence from the loss of connectivity of the atoms or molecules due to thermal expansion, where the thermal-expansion parameter $\alpha_T$ is determined by the anharmonic part of the attractive potential and a power exponent $\lambda$ covers the repulsive part of the potential[32,36] The latter is also termed steepness parameter, describing the strong increase of the potential due to the Born-Mayer electron overlap repulsion. The predictive power of this model was tested for a series of La-based metallic melts.[37] In this KSZ model, the shear modulus $G$ depends exponentially on temperature via a prefactor $(2 + \lambda)\,\alpha_T$. An exponential decay of $G$ vs $T$ has been shown by ultrasound measurements and, even more convincingly, by molecular-dynamics simulations.[38] Since the viscosity $\eta$ itself depends exponentially on $G$, the viscosity becomes a double-exponential function on $T$ as

$$\eta(T) = \eta_0 \exp\left\{\frac{V_c C_G}{k_\mathrm{B} T}\exp\left[(2+\lambda)\alpha_T T_g\left(1-\frac{T}{T_g}\right)\right]\right\} \qquad (1).$$

Here $\eta_0$ is a normalization constant, $V_c$ is the activation or shear-transformation zone volume, $C_G$ is a temperature independent prefactor, $k_\mathrm{B}$ is the Boltzmann constant, and $T_g$ the glass-transition temperature (see Ref. 32 for more details). Equation (1) essentially is a three-parameter function (with an additional parameter $T_g$ taken from literature). We want to remark that this approach is not just another fitting function of the super-Arrhenius behavior of the vitrification process, but it is a model prediction using well-defined microscopic parameters. One should note that Eq. (1) can also be formulated without including $T_g$,[32] which is not a precisely defined material property but somewhat depends on its definition and measurement method. For this purpose, one can use equations (4) and (5) of Ref. 32, both providing expressions for the high-frequency shear modulus $G(T)$, without or with $T_g$, respectively, the latter being contained in our eq. (1). This leads to: $\eta(T) = \eta_0 \exp\{c/T \exp[-(2+\lambda)\alpha_T T]\}$ with the fit-parameters $\eta_0$, $c$, and $(2+\lambda)\alpha_T$. Here $c$ is the product of several temperature-independent model parameters that cannot be independently determined in the fits. In any case, it should be noted that, even in the original model prediction, eq. (1), the most relevant fit parameter $\lambda$ (or $(2+\lambda)\alpha_T$ if $\alpha_T$ is unknown) does not depend on the choice of $T_g$ because, essentially, the $T_g$-dependent part of the argument of the second exponential simply represents a prefactor. Moreover, fits with eq. (1) or the $T_g$-free version of the equation, given above, lead to the same $\lambda$.

It is well known that the non-Arrhenius temperature dependence of glassy dynamics is a universal property of all glass-forming matter. Thus one may ask whether the KSZ model also is applicable to non-metallic glass formers. Here we present an extension of the KSZ model to describe, in addition to metallic glass-forming liquids, a variety of strong and fragile glass formers, including hydrogen and van-der-Waals bonded molecular systems, ionic liquids, covalent-network systems, polymers, and plastic crystals. We systematically modelled 31 systems with the approach of the KSZ model and fitted the mean dipolar relaxation time $\tau$ of most of these systems. To elucidate the predictive power of the model and the correlations of the main parameters, in the following we discuss in detail several representative systems. Further results for all systems analyzed in the course of this work are shown in the supplementary material (SM). We find that the KSZ model indeed allows describing very different glass-forming systems. We explore the relation of the main fitting parameter $(2 + \lambda)\,\alpha_T$ (and of $\lambda$, for systems where $\alpha_T$ is known) to the fragility $m$.

## II. ANALYSIS AND DISCUSSION

In the following, we will mainly compare the temperature dependence of the mean relaxation time $\tau$ as previously determined by dielectric spectroscopy performed by the Augsburg group, in many cases over more than 10 decades of time, with the elastic model outlined above. A brief summary of the dielectric measuring techniques used to obtain these broadband data for frequencies ranging from the mHz up to the GHz regime can be found in Refs. 11 and 39.

### A. Comparison of the analysis of broadband experiments using the VFT, the Mauro, and the KSZ model.

Below, we will present the temperature dependence of mean relaxation times as determined by dielectric spectroscopy of four out of the large variety of investigated systems, belonging to very different classes of disordered matter. Information on all the materials investigated or analyzed in the course of this work is documented in the SM. Figure 1(a) represents an Arrhenius plot of the mean relaxation times of glycerol,[15] which is a prime example of low-molecular-weight glass-forming liquids and one of the most thoroughly investigated glass-forming system. Representative spectra of the real and imaginary parts of the dielectric permittivity covering more than 18 decades in frequency can be found in Ref. 29. From these and similar results, the temperature dependence of $\tau$ was extracted by simultaneous fits of the real and imaginary part. In addition, results from aging experiments at $T < T_g$ (Ref. 40) were included in the final relaxation map which extends over 16 decades of time.

A critical comparison of fits of broadband relaxation-time data using the Mauro formula and the VFT law was reported earlier by some of the authors.[15] In the following, we in addition provide fits of an extended data basis with the newly proposed KSZ model and compare the outcome with the two other approaches. The best fits of the glycerol data, utilizing the VFT law, the Mauro equation, and the KSZ model, are shown by the lines in Fig. 1(a). The KSZ model provides an almost as good fit as the VFT and Mauro equation, with slight deviations at low and very high temperatures. The parameter values for $(2 + \lambda)\,\alpha_T$ obtained for this and the other materials are given in Table S1 of the SM. During the fits with Eq. (1), for glycerol and the other systems the glass temperature $T_g$ always was fixed to the known literature values (Table S1). As shown in detail in the SM, in a similar way we also analyzed the dielectric relaxation times of various other molecular glass formers built by polar molecules, including both, hydrogen and van-der-Waals bonded systems,[15,40,41,42,43,44,45,46] essentially leading to the same result: the KSZ model is able to provide a reasonable description of the typical super-Arrhenius behavior of all these glass formers.

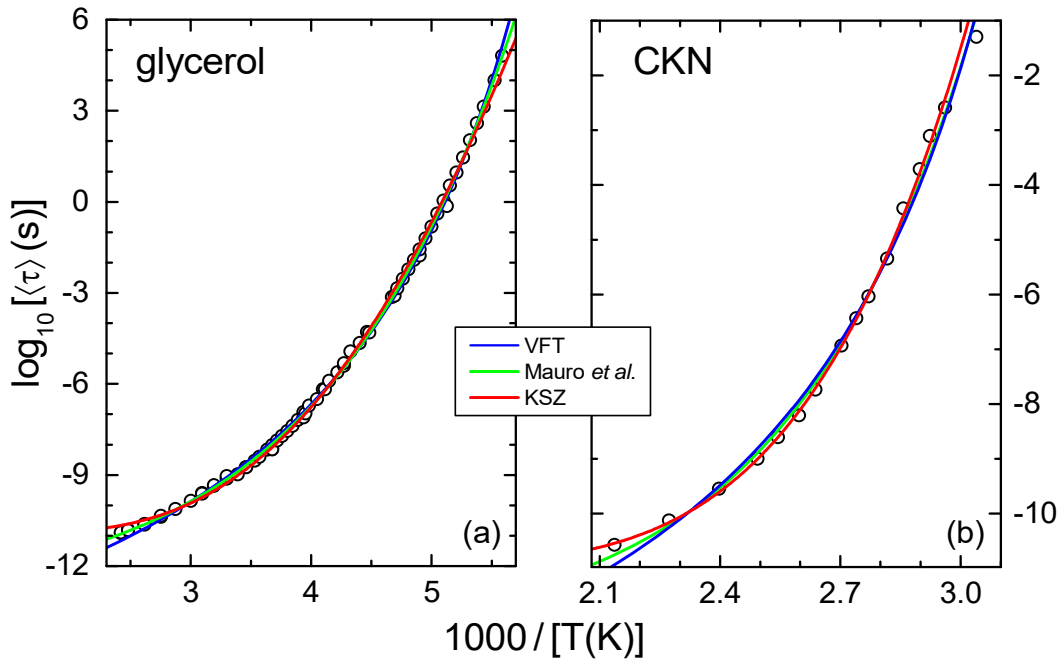

FIG. 1. Arrhenius representation of the mean relaxation times as determined via dielectric spectroscopy. (a) Results of the low-molecular-weight supercooled-liquid glycerol. The experimental data were taken from Ref. 15. (b) Results as determined in the ionic conductor CKN. Here the mean relaxation times were determined from fits of the modulus function.[43] The lines show fits of the experimental results using the VFT, the Mauro, as well as the KSZ model.

Figure 1(b) shows the relaxation time as a function of the inverse temperature for the ionic melt [Ca(NO$_3$)$_2$]$_{0.4}$[KNO$_3$]$_{0.6}$ (CKN). In the case of this ionically conducting molten salt, the mean relaxation times were determined from an analysis of the modulus function,[43,47] which corresponds to the inverse of the complex permittivity.[48] Again, this glass former reveals a clear super-Arrhenius behavior extending over almost 10 time decades. As documented in Fig. 1(b), the mean relaxation times of CKN are best described by the KSZ model, which even matches the high-temperature behavior quite reasonably and significantly better than the empirical VFT equation or the Mauro entropy model. A corresponding analysis of broadband relaxation-time data of several other ionically conducting glass formers,[49,50,51] including various ionic liquids[50] and a deep eutectic solvent,[51,52] also revealed good fits by the KSZ model (see SM for details). In the latter two material classes, in contrast to CKN, reorientational motions of dipolar entities are possible, enabling the analysis of the dielectric permittivity instead of the modulus (see Refs. 50 and 52 for details).

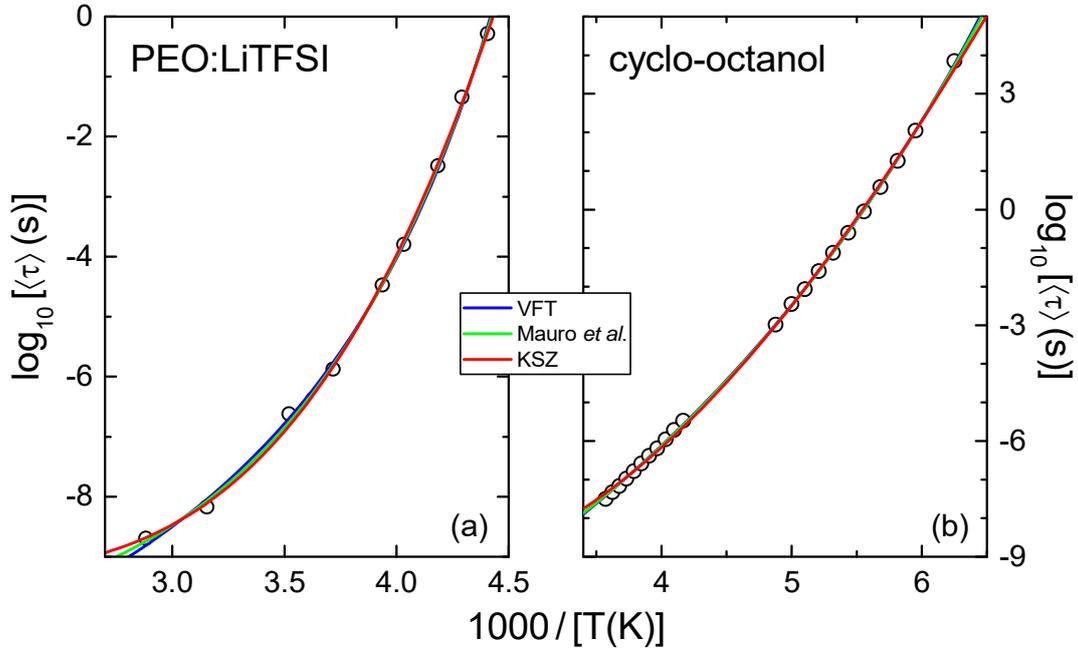

FIG. 2. Mean relaxation times as determined via dielectric spectroscopy of the polymer PEO:LiTFSI[53] (a) and of the plastic crystal cyclo-octanol (b).[54] The super-Arrhenius behavior in both compounds is fitted by the VFT law, the Mauro model, as well as the KSZ model.

Figure 2(a) presents the mean $\alpha$-relaxation times as determined via dielectric spectroscopy for a polymeric system doped with a lithium salt, taken from Do et al.[53] Due to its excellent ion-transport ability, polyethylene oxide (PEO) is one of the most widely studied polymers for the future use as solid electrolyte in lithium-ion batteries. In the example documented in Fig. 2(a), the polymer was doped with lithium bis(trifluoromethanesulfonyl)imde (LiTFSI). Here the temperature-dependent relaxation times span nine decades. The found super-Arrhenius behavior again was analyzed in terms of the VFT, Mauro, and KSZ formulae as indicated by the lines in Fig. 2(a). The KSZ model obviously provides a convincing fit of the experimental data, which is of similar quality as for the other approaches.

Figure 2(b) shows the relaxation dynamics as determined in a plastic crystal.[54] Plastic crystals are a specific class of disordered matter, with long-range order of the center-of-mass lattice, formed by asymmetric molecules, but with dynamically disordered orientational degrees of freedom. The freezing process of the latter often resembles the cooperative behavior also found for dipolar supercooled liquids and exhibits super-Arrhenius behavior.[55] Hence, plastic crystals are often considered as model systems for glassy freezing. Figure 2(b) presents results for the temperature dependence of the mean relaxation times as determined in plastic-crystalline cyclo-octanol,[54] covering more than 11 decades of time, which again can be reasonably well fitted by the KSZ model, Eq. (1). Similar findings were obtained for $\tau(T)$ of three other plastic crystals[55,56,57] as shown in the SM.

Finally, we discuss the case of the prototypical glass-forming oxide silica ($SiO_2$), which, according to Angell's classification scheme,[5] belongs to the strongest of all glass-forming materials ($m = 20$; Ref. 7) like most of the inorganic tetrahedrally coordinated network liquids. Dielectric spectroscopy cannot provide information on the dynamics in this material, due to the lack of dipolar moments or mobile ions. Thus, we analyzed its viscosity, taken from the literature,[5,58,59] which rather closely follows

Arrhenius behavior. As documented in Ref. 5, the Arrhenius plot of $\eta$ covers about 9 decades as reproduced in Fig. 3. In this plot, the dash-dotted line indicates the maximum strong behavior, using the common value of $10^{13}$ P for the viscosity at $T_g$ (1446 K in this case5) and the approach of a value of $10^{-4}$ P for $T \to \infty$ as assumed in Fig. 3 of Ref. 5. Small but significant deviations from this ideal strong temperature dependence are clearly identified. As shown by the solid lines in Fig. 3, the temperature-dependent viscosity can be equally well described utilizing the VFT and Mauro formulas or the KSZ model (due to the sparse data base, during the fits we fixed $\eta_0$ to a value of $10^{-4}$ P as suggested by the Angell plot, Fig. 3 of Ref. 5). We also analyzed the viscosity results from literature for three more strong glass formers,[60,61,62,63] also leading to a good description by Eq. 1 (see SM).

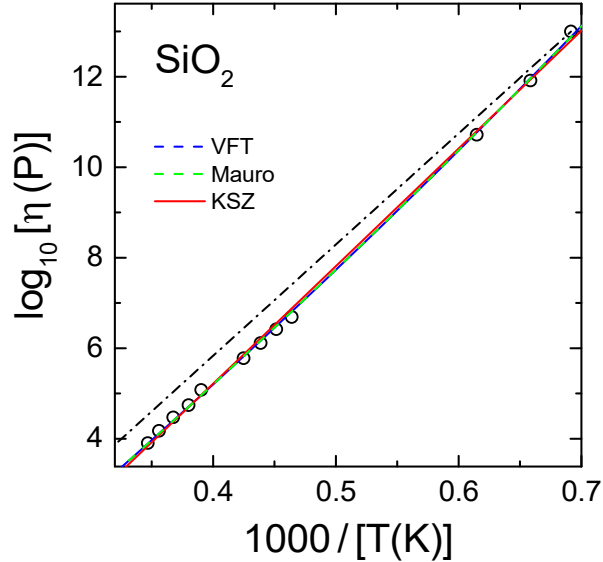

FIG. 3. Arrhenius plot of the viscosity of $SiO_2$. The viscosity data are reproduced from Ref. 5 and fitted using the KSZ, VFT, and Mauro equations. In all cases, $\eta_0 = \eta(T \to \infty)$ was fixed to $10^{-4}$ P.[5] The dash-dotted line indicates ideal Arrhenius behavior assuming $\eta(T_g) = 10^{13}$ P and $\eta_0 = 10^{-4}$ P.[5]

In all cases presented here and in the SM, we found reasonable agreement of the experimental results of the temperature dependence of the mean relaxation times or viscosities with the KSZ model prediction, Eq. (1). This is true, despite the fact that the glass-forming liquids investigated in the course of this work cover very different material classes, including hydrogen and van-der-Waals bonded molecular materials, covalent-network glass formers, ionic melts and liquids, polymers, and even plastic crystals. One should be aware that sometimes it is assumed that $\tau(T)$ or $\eta(T)$ show a crossover from super-Arrhenius into pure Arrhenius behavior in the low-viscosity regime at high temperatures.[64,65] Indeed, there are several models that explicitly involve such a crossover[20,22,23,24,25] which were applied to describe experimental data, e.g., in Refs. 24,25,29,30,39,41. However, in the above analysis the KSZ model, not assuming such a transition, is tested and compared with other models within the complete dynamic range from $T_g$ well into the low-viscosity liquid. From our point of view this is an unbiased approach. In what follows, we will present and discuss scaling and correlations of the $\lambda$ parameter with $m$ for all these liquids with very different types of bonding and covering the complete range from very strong to very fragile glass formers.

**B. Correlations of steepness parameter $\lambda$ and thermal-expansion parameter $\alpha_T$**

As mentioned in the beginning, the KSZ model, which goes back to the Born-Huang-ansatz and the shoving-model, significantly depends on two parameters, mainly representing the repulsive and attractive parts of the molecular pair potential. Starting from this ansatz, the connectivity is modelled and concomitantly the shear modulus evaluated. Using the shear modulus as the dominant thermodynamic variable, the viscosity can be computed resulting in a double-exponential functional (Eq. 1). The prefactor of the second exponential function is determined by $(2 + \lambda)\, \alpha_T$ times the glass-transition temperature. While $T_g$ is known for all investigated systems, this is not the case for $\alpha_T$. Therefore, $(2 + \lambda)\, \alpha_T$ was used as free fitting parameter. In Fig. 4(a), we check for a possible correlation of this parameter with the fragility by plotting it versus $m$ for all materials analyzed in the present work (a list of all materials and their parameters is given in Table S1 of the SM). In addition, we also include results for four metallic glass formers with known $m$, taken from Ref. 32. For most systems, an overall linear dependence of the parameter $(2 + \lambda)\, \alpha_T$ as function of fragility can be identified. The most fragile glass-forming liquids are represented by compounds with predominantly ionic type of bonding, where $m$ exhibits values up to 150. For these systems, we get a value of 0.03 for the combined parameter $(2 + \lambda)\, \alpha_T$.

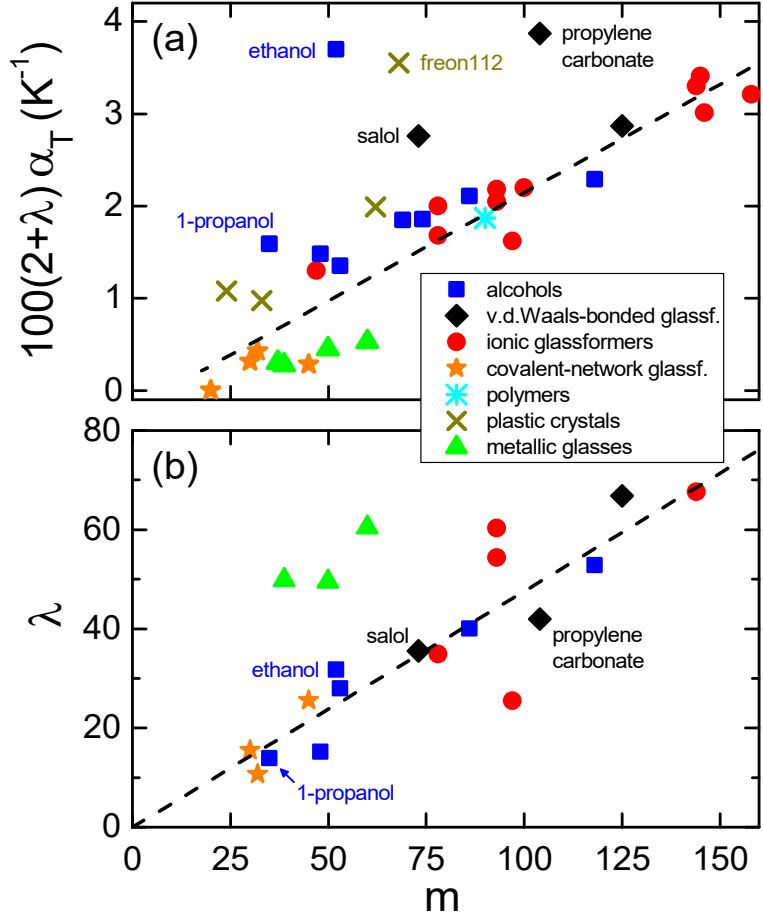

FIG. 4. Correlations of steepness parameter $\lambda$ and thermal-expansion parameter $\alpha_T$ for the variety of different non-metallic glass formers analyzed in the course of this work (see Table S1 in the SM for a complete list). In addition, data for four metallic glass formers from Ref. 32 are included.[66] (a) Plot of the fit parameter $(2 + \lambda)\, \alpha_T$ versus the fragility index $m$. The dashed line indicates an approximately linear correlation of the two quantities. Materials that strongly deviate from this correlation are labeled in the figure. (b) Correlation of the steepness parameter $\lambda$ and $m$ for the systems with known $\alpha_T$. The dashed line is a fit of all non-metal data points, assuming direct proportionality $\lambda = s \cdot m$, which leads to $s \approx 0.48$.

One should be aware that the thermal-expansion coefficients, related to the anharmonic attractive part of the pair potential, strongly varies from alcohols (of order $10^{-3}\,\mathrm{K}^{-1}$) over covalent systems like $Na_2Si_2O_5$ (around $10^{-4}\,\mathrm{K}^{-1}$) to metallic glasses (typically $10^{-5}\,\mathrm{K}^{-1}$). These values span roughly two decades and demonstrate the enormous variety of non-harmonic interactions. This may partly explain the considerable deviations from the suggested correlation of the combined parameter $(2 + \lambda)\, \alpha_T$ and $m$, observed for several glass formers in Fig. 4(a). Therefore, we make use of the experimental data basis known for the thermal-expansion coefficient (see Table S1 in SM) to calculate $\lambda$. Here we used $\alpha_T$ in the liquid region, as the model is applied for temperatures $T > T_g$. The obtained $\lambda$ values are plotted in Fig. 4(b) *vs.* the fragility index. The thermal-expansion values are not known for all systems represented in Fig. 4(a) and, hence, the data basis is somewhat reduced. However, Fig. 4(b), where $\lambda$ is plotted *vs. m*, still is a representative plot for most of the different categories of glass-forming liquids. For all materials, except the metallic glasses, we find a significant correlation of both quantities with much reduced scatter compared to frame (a). Obviously, for increasing *m* the steepness parameter of the repulsive part of the pair potential increases significantly. This correlation now also includes glass formers like ethanol or propylene carbonate that revealed strong deviations in Fig. 4(a). On the other hand, the $\lambda$ values of the metallic glass formers, deduced[66] from the fits in Ref. 32, clearly deviate from the universal correlation of $\lambda$ and *m* found for the other materials. The dashed line in Fig. 4(b) is a linear fit of all non-metal data points, suggesting a direct proportionality, $\lambda = s \cdot m$ with $s \approx 0.48$. Notably, the KSZ model also predicts a linear relation between $\lambda$ and *m* [cf. Eq. (7) in Ref. 32], however, with an intercept $\lambda(m=0)$ below zero. This intercept can be estimated, based on the values of $\alpha_T$ and $T_g$ in Table S1 of the SM, to be of the order of -12, which is compatible with the experimental data in Fig. 4(b).

Within the KSZ model, $\lambda$ represents the power-law exponent used to describe the repulsive part of the interaction potential. Power-law exponents of 50-60 or more, at first glance may seem hard to understand concerning their physical meaning. Thus a power law only may be regarded as a formal description of the potential, which certainly deserves further theoretical consideration in the future.

### III. SUMMARY AND CONCLUSIONS

In summary, we have demonstrated that the use of a purely mechanical ansatz based on the Born-Huang-equation for the shear modulus and using the shoving model to calculate the viscosity, which so far has been applied to metallic glasses only,[32,36,37] can be successfully expanded to a large variety of non-metallic glass formers, including molecular supercooled liquids like glycerol, liquid salts, polymers, covalent-network systems, and even plastic crystals. In most of these materials, we used the mean relaxation times instead of the viscosities for the analysis. Indeed, the scaling of dipolar relaxation times with viscosity data has been proven experimentally for a variety of glass-forming liquids.[13,14] Here we have employed this ansatz for the analytical description of the temperature-dependent change in $\tau$ or $\eta$ over more than 10 decades of time or viscosity. The parameters used in this model are based on a steepness parameter $\lambda$, characterizing the strength of the repulsive part of the interaction potential, and on the physical quantity $\alpha_T$, the thermal expansion coefficient related to the anharmonic attractive part. In the present work, we demonstrate for a variety of glassy systems that this microscopic ansatz provides an at least as good fit of the super-Arrhenius behavior, as can be achieved utilizing the phenomenological VFT law or the Mauro equation, recently proposed as an alternative to the VFT equation.

In addition to the fact that this microscopic ansatz provides reasonable fits to the viscosities or the mean relaxation times of a large number of glassy systems, this model predicts a linear dependence of the fragility *m* on the steepness of the interatomic potential $\lambda$.[32] Using the values of $\lambda$ as obtained from the fits of the experimental data and the fragilities as listed in Table S1, we indeed find such a linear dependence, well compatible with the model predictions. Thus, the microscopic model proposed by Krausser, Samwer, and Zaccone,[32] not only describes the super-Arrhenius behavior of a variety of glass formers, but also hints to a universal dependence of the fragility on the repulsive part of the intermolecular potential. The fact that this model is based on microscopic model considerations and in some respect "explains" fragility by tracing it back to microscopic model parameters, makes it stand out from most of the numerous other alternatives to the empirical VFT approach that were proposed during the past decades.

**SUPPLEMENTARY MATERIAL**

See the supplementary material for relaxation-time plots with fits for all systems analyzed in the present work and for a table with a list of all materials and their parameters.

**ACKNOWLEDGMENTS**

K.S. is thankful for numerous discussions with A. Zaccone, A. Lagoggiani, and Z. Wang. K.S. also received support from the DFG via the Leibniz-Programm and the SFB 1073.

**DATA AVAILABILITY**

The data that support the findings of this study are available within the article and its supplementary material.

# Supplementary Material
for
# Universal Correlations Between the Fragility and Interparticle Repulsion of Glass-Forming Liquids

Peter Lunkenheimer[1], Felix Humann[1], Alois Loidl[1], and Konrad Samwer[2]
[1]Experimental Physics V, Center for Electronic Correlations and Magnetism, University of Augsburg, 86159 Augsburg, Germany
[2]I. Physikalisches Institut, Universität Göttingen, 37077 Göttingen, Germany

| material | $T_g$ (K) | $m$ | $100(2+\lambda)\alpha_T$ (K$^{-1}$) | $10^4 \alpha_T$ (K$^{-1}$) | $\lambda$ |
|---|---|---|---|---|---|
| **alcohols** | | | | | |
| ethanol | 99 [1] | 52 [1] | 3.7 | 11.0 [2] | 31.8 |
| 1-propanol | 96 [3] | 35 [4] | 1.59 | 10.0 [5] | 13.9 |
| propylene glycol | 168 [1] | 48 [6] | 1.48 | 8.59 [7] | 15.2 |
| dipropylene glycol | 193 [1] | 69 [6] | 1.85 | | |
| tripropylene glycol | 189 [1] | 74 [6] | 1.86 | | |
| glycerol | 185 [1] | 53 [8] | 1.35 | 4.5 [9] | 28.0 |
| xylitol | 248 [1] | 86 [1] | 2.11 | 5.02 [10] | 40.0 |
| sorbitol | 274 [8] | 118 [1] | 2.29 | 4.18 [11] | 52.8 |
| **van-der-Waals glass formers** | | | | | |
| propylene carbonate | 159 [1] | 104 [8] | 3.87 | 8.8 [12] | 42.0 |
| salol | 218 [8] | 73 [8] | 2.76 | 7.36 [13] | 35.5 |
| benzophenone | 212 [14] | 125 [15] | 2.87 | 4.17 [16] | 66.8 |
| **ionic glassformers** | | | | | |
| [Ca(NO$_3$)$_2$]$_{0.4}$[KNO$_3$]$_{0.6}$ (CKN) | 333 [17] | 93 [8] | 2.05 | 3.64 [18] | 54.3 |
| [Ca(NO3)$_2$]$_{0.4}$[RbNO$_3$]$_{0.6}$ (CRN) | 333 [17] | 100 [1] | 2.20 | | |
| 1-Butyl-3-methylimidazolium tetrafluoroborat (Bmim BF$_4$) | 182 [19] | 93 [19] | 2.18 | 3.5 [20] | 60.3 |
| 1-Butyl-3-methylimidazolium chloride (Bmim Cl) | 228 [19] | 97 [19] | 1.62 | 5.9 [21] | 25.5 |
| 1-Butyl-3-methylimidazolium tetrachloroferrate (Bmim FeCl$_4$) | 182 [19] | 144 [19] | 3.30 | 4.74 [22] | 67.6 |
| 1-Butyl-3-methylimidazolium bromotrichloroferrate (Bmim FeCl$_3$Br) | 180 [19] | 146 [19] | 3.01 | | |
| 1-Benzyl-3-methyl-imidazolium chlorid (Benzmim Cl) | 253 [19] | 78 [19] | 1.68 | | |
| 1-Ethyl-3-methyl-imidazolium tricyanomethanide (Emim TCM) | 183 [19] | 158 [19] | 3.21 | | |
| 1-Methyl-3-octylimidazolium hexafluorophosphate (Omim PF$_6$) | 194 [19] | 78 [19] | 2.00 | 5.42 [23] | 34.9 |
| 1,3-Dimethylimidazolium(Li 1.0m) bis-(trifluoromethylsulfonyl)imide ([Li+Dimim]TFSI) | 202 [19] | 145 [19] | 3.41 | | |
| glyceline | 175 [24] | 47 [24] | 1.30 | | |
| **covalent network glassformers** | | | | | |
| SiO$_2$ | 1446 [25] | 20 [8] | 0.00503 | | |
| Na$_2$Si$_2$O$_5$ | 713 [25] | 45 [8] | 0.284 | 1.03 [26] | 25.6 |
| B$_2$O$_3$ | 554 [8] | 32 [8] | 0.426 | 3.35 [27] | 10.7 |
| ZnCl$_2$ | 380 [8] | 30 [8] | 0.316 | 1.8 [28] | 15.5 |
| **polymer** | | | | | |
| (polyethylene oxide):lithium-bis(trifluoromethanesulfonyl)imde (PEO:LiTFSI) | 219 [29] | 90 [29] | 1.87 | | |
| **plastic crystals** | | | | | |
| cyanoadamantane | 178 [30] | 24 [30] | 1.08 | | |
| cyclo-octanol | 168 [31] | 33 [31] | 0.978 | | |
| Freon112 | 88 [32] | 68 [32] | 3.55 | | |
| (succinonitrile)$_{0.6}$(glutaronitrile)$_{0.4}$ (SN$_{0.6}$GN$_{0.4}$) | 144 [33] | 62 [33] | 1.99 | | |
| **metallic glass formers** | | | | | |
| La$_{55}$Al$_{25}$Ni$_{20}$ | 465 [34] | 37 [35] | 0.303 [35] | | |
| Zr$_{41.2}$Ti$_{13.8}$Ni$_{10}$Cu$_{12.5}$Be$_{22.5}$ | 623 [34] | 39 [35] | 0.276 [35] | 0.532 [36] | 49.8 |
| Pd$_{40}$Ni$_{40}$P$_{20}$ | 551 [34] | 50 [35] | 0.448 [35] | 0.87 [37] | 49.5 |
| Pd$_{77.5}$Cu$_6$Si$_{16.5}$ | 625 [34] | 60 [35] | 0.525 [35] | 0.84 [38] | 60.5 |

TABLE S1. Glass temperature $T_g$, fragility index $m$, fit parameter $(2+\lambda)\alpha_T$, thermal volume expansion coefficient $\alpha_T$ in the liquid state, and exponent $\lambda$ for all investigated glass formers (the numbers in brackets refer to the reference list given at the end of this document). Values for $\alpha_T$ (and thus for $\lambda$) are not available for all systems. For the metallic glass formers, $(2+\lambda)\alpha_T$ was taken from Ref. 35 and $\lambda$ was recalculated using the volume expansion coefficient above $T_g$.

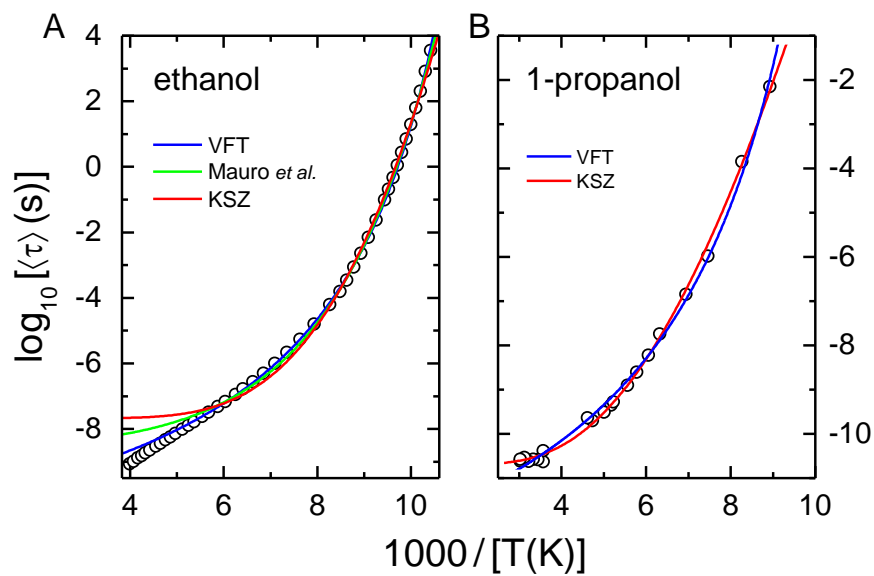

FIG. S1. Arrhenius representation of the mean $\alpha$-relaxation times as determined via dielectric spectroscopy for the alcohols ethanol (A) and 1-propanol (B). The experimental data were taken from Refs. 1 and 39, respectively. The lines show fits using the VFT, the Mauro (only for ethanol [1]), and the KSZ model. For ethanol, only the data at $T < 190$ K (i.e., $1000/T > 5.3$ K$^{-1}$) were fitted.

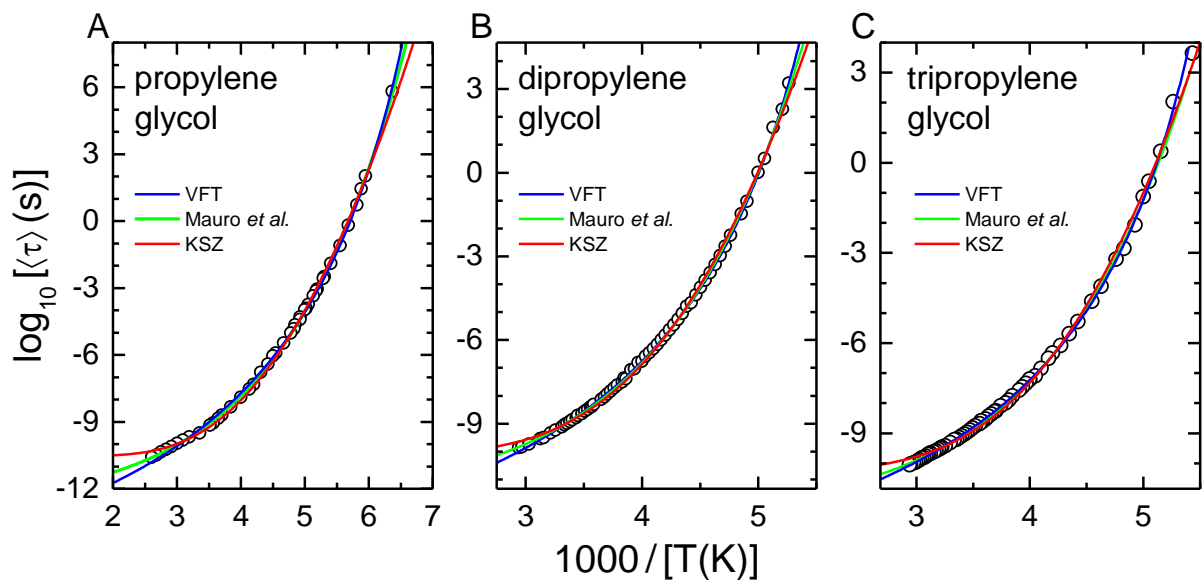

FIG. S2. Arrhenius representation of the mean $\alpha$-relaxation times as determined via dielectric spectroscopy for the alcohols propylene glycol (A), dipropylene glycol (B), and tripropylene glycol (C). The experimental data were taken from Ref. 1. The lines show fits using the VFT, the Mauro, and the KSZ model.

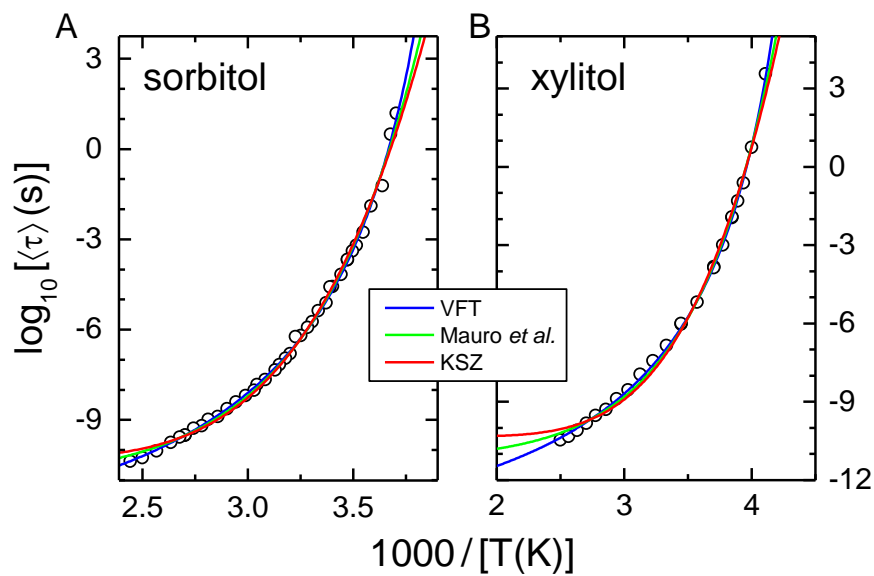

FIG. S3. Arrhenius representation of the mean α-relaxation times as determined via dielectric spectroscopy for the alcohols sorbitol (A) and xylitol (B). The experimental data were taken from Ref. 1. The lines show fits using the VFT, the Mauro, and the KSZ model.

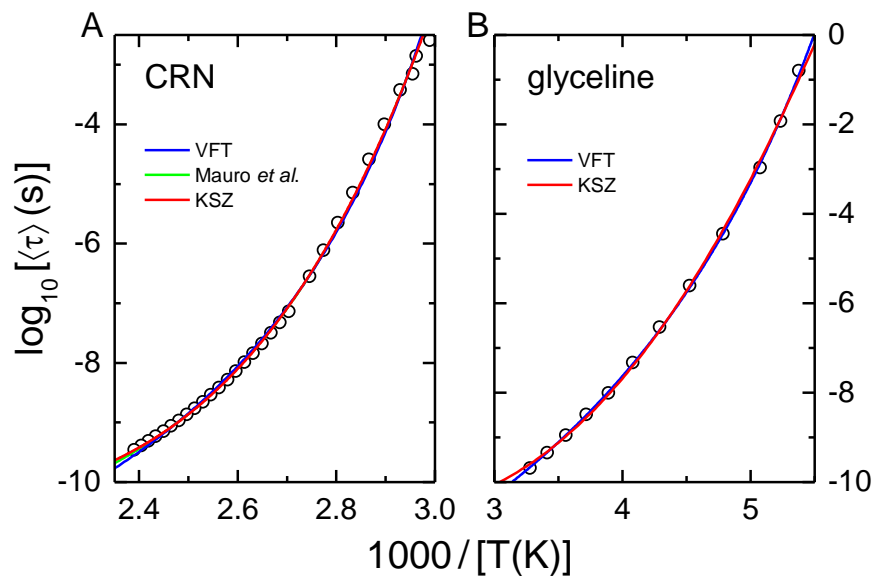

FIG. S4. Arrhenius representation of the mean α-relaxation times as determined via dielectric spectroscopy for the ionic melt $[Ca(NO_3)_2]_{0.4}[RbNO_3]_{0.6}$ (CRN) (A) and for the deep eutectic solvent glyceline (B). The experimental data were taken from Refs. 17 and 40, respectively. The lines show fits using the VFT, the Mauro (only for CRN[1]), and the KSZ model.

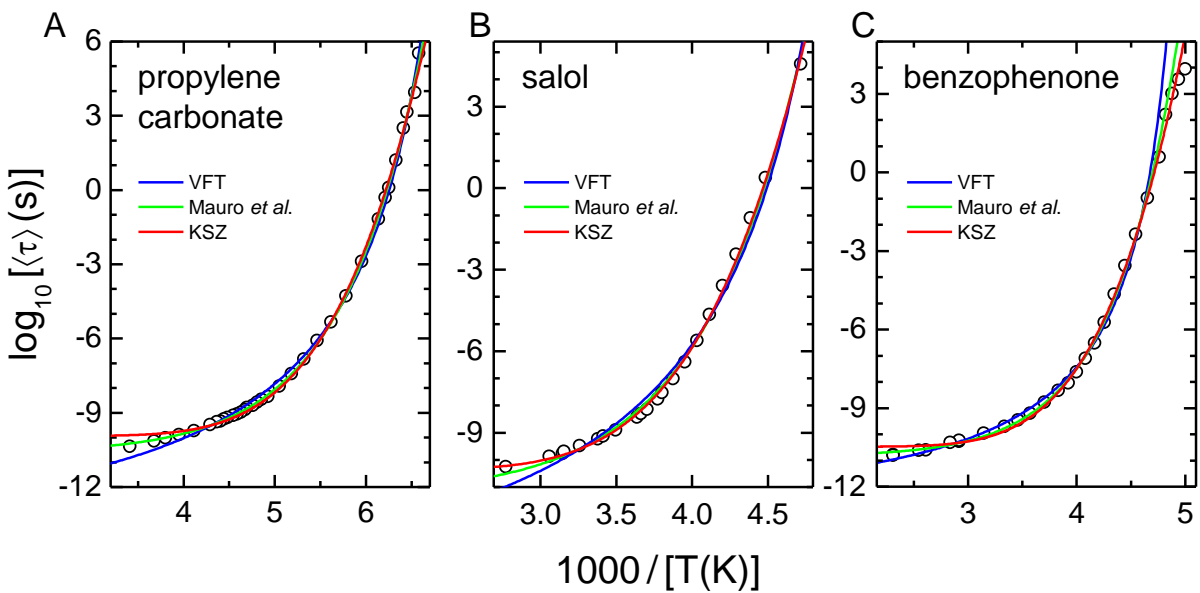

FIG. S5. Arrhenius representation of the mean α-relaxation times as determined via dielectric spectroscopy for the van-der-Waals-bonded molecular glass formers propylene carbonate (A), salol (B), and benzophenone (C). The experimental data were taken from Ref. 1. The lines show fits using the VFT, the Mauro, and the KSZ model.

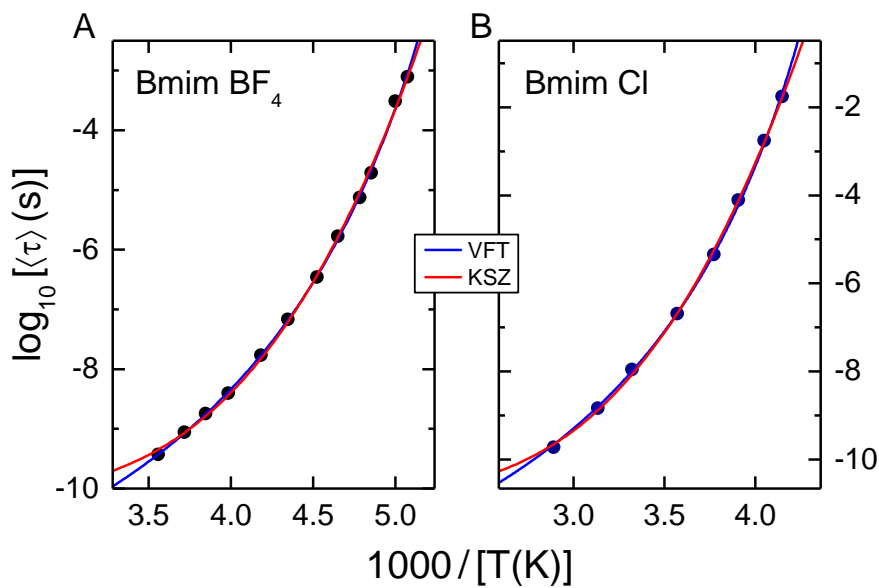

FIG. S6. Arrhenius representation of the mean α-relaxation times as determined via dielectric spectroscopy for the ionic liquids Bmim $BF_4$ (A) and Bmim Cl (B). The experimental data were taken from Ref. 19 and deduced via an analysis of the dielectric modulus. The lines show fits using the VFT and the KSZ model.

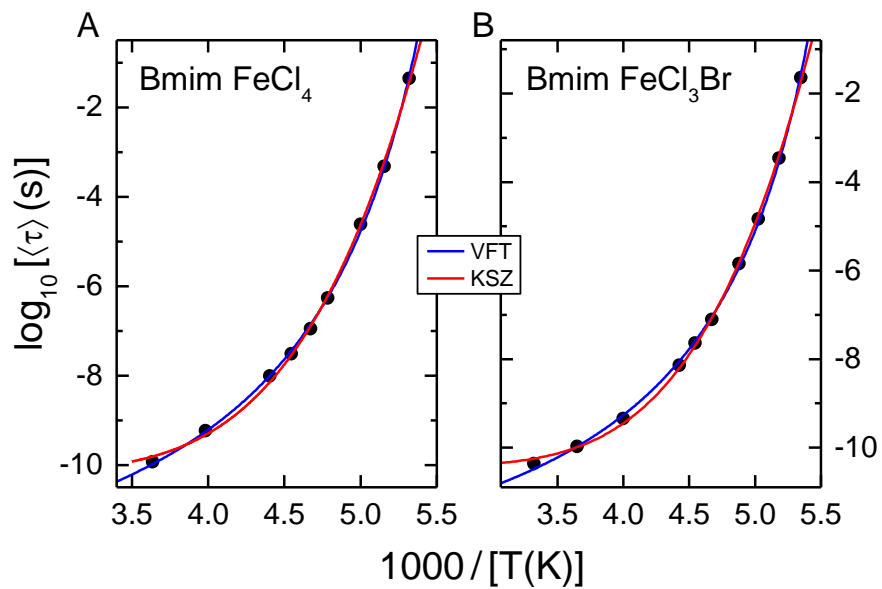

FIG. S7. Arrhenius representation of the mean α-relaxation times as determined via dielectric spectroscopy for the ionic liquids Bmim FeCl$_4$ (A) and Bmim FeCl$_3$Br (B). The experimental data were taken from Ref. 19 and deduced via an analysis of the dielectric modulus. The lines show fits using the VFT and the KSZ model.

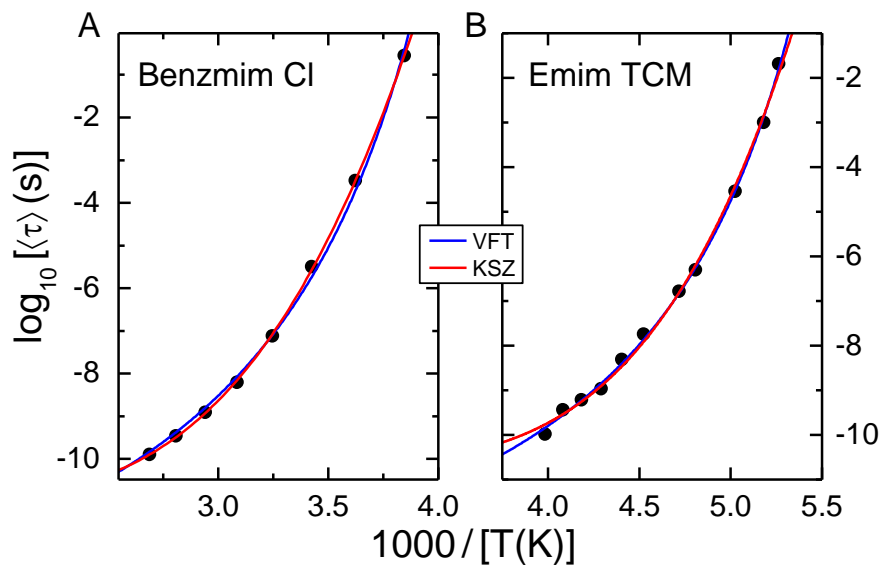

FIG. S8. Arrhenius representation of the mean α-relaxation times as determined via dielectric spectroscopy for the ionic liquids Benzmin Cl (A) and Emim TCM (B). The experimental data were taken from Ref. 19 and deduced via an analysis of the dielectric modulus. The lines show fits using the VFT and the KSZ model.

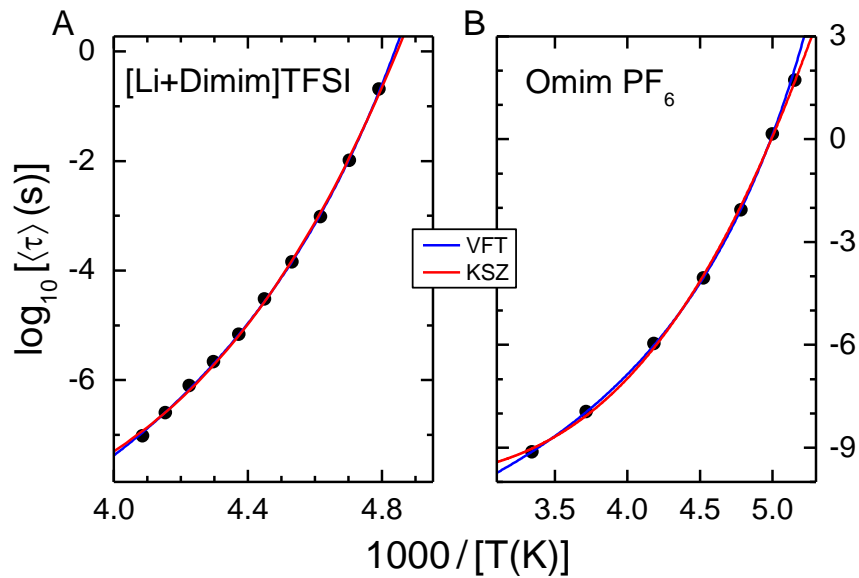

FIG. S9. Arrhenius representation of the mean $\alpha$-relaxation times as determined via dielectric spectroscopy for the ionic liquids [Li+Dmim]TFSI (A) and Omim PF$_6$ (B). The experimental data were taken from Ref. 19 and deduced via an analysis of the dielectric modulus. The lines show fits using the VFT and the KSZ model.

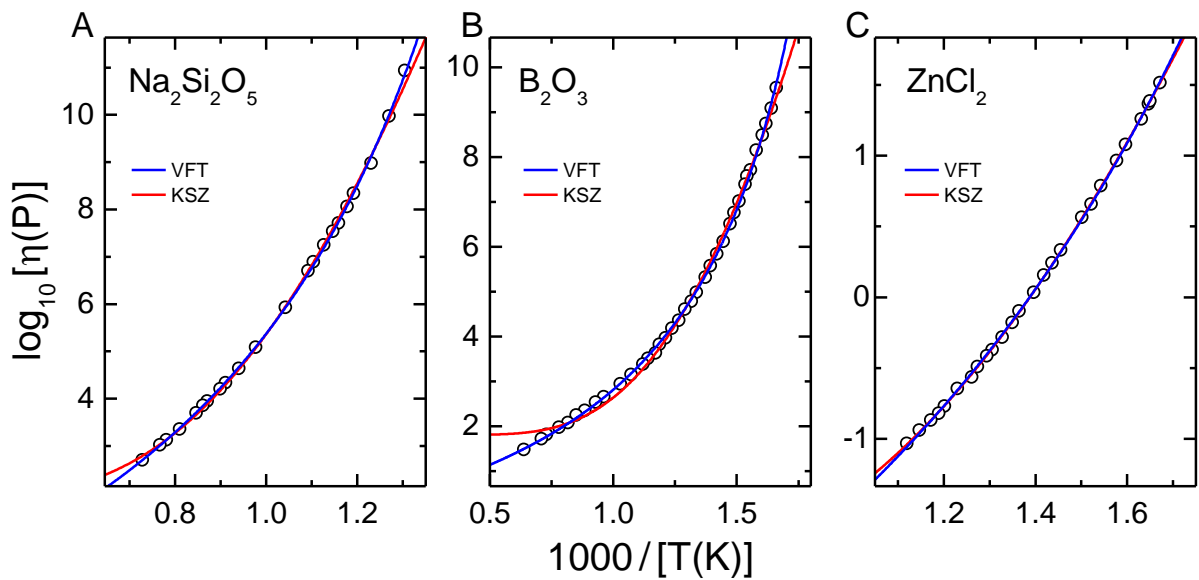

FIG. S10. Arrhenius representation of the viscosity for the covalent network glass formers Na$_2$Si$_2$O$_5$ (A), B$_2$O$_3$ (B), and ZnCl$_2$ (C). The experimental data in (A) and (B) were taken from Refs. 41 and 27, respectively. The data in (C) are from Ref. 42 (for $T < 725$ K) and Ref. 43 (for $T > 725$ K). The lines show fits using the VFT and the KSZ model.

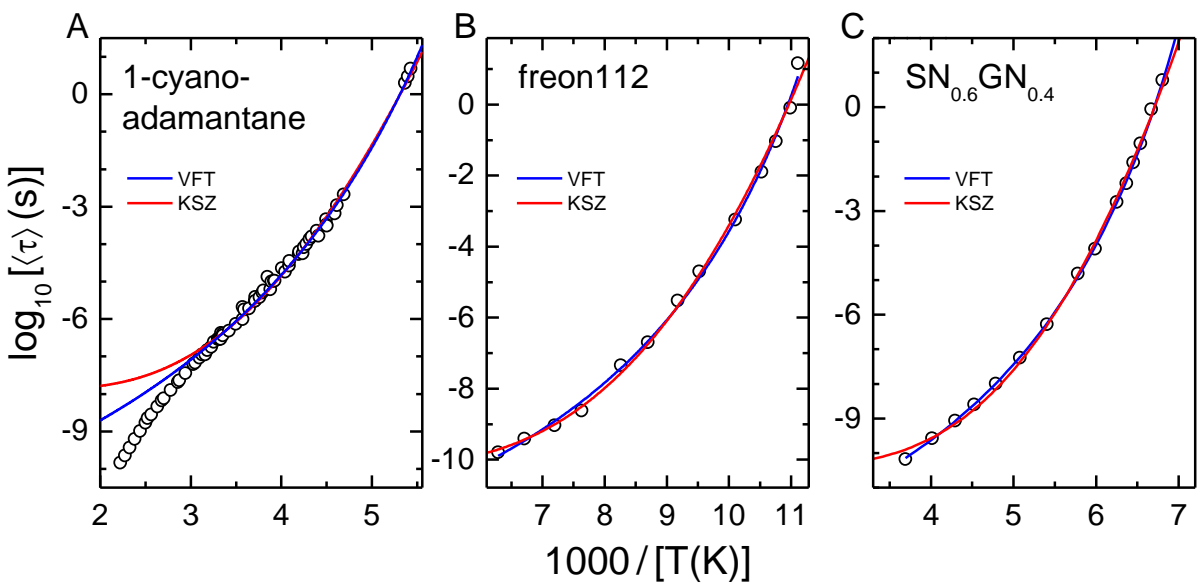

FIG. S11. Arrhenius representation of the viscosity for the covalent plastic crystals 1-cyano-adamantane (A), freon112 (B), and $SN_{0.6}GN_{0.4}$ (C). The experimental were taken from Refs. 30, 32] and 33, respectively. The lines show fits using the VFT and the KSZ model. For 1-cyano-adamantane, only data points at $T < 315$ K (i.e., $1000/T > 3.2$ K$^{-1}$) were used for the fits.